\documentclass[runningheads]{llncs}
\usepackage[utf8]{inputenc}

\usepackage[numbers]{natbib}
\usepackage{graphicx}
\usepackage{longtable}
\usepackage{caption}
\usepackage{comment}
\usepackage{subcaption}
\usepackage{url}

\begin{document}

\title{Detecting Anti-vaccine Content on Twitter using Multiple Message-Based Network Representations}

%
%

\author{James R. Ashford \inst{1}\orcidID{0000-0002-2678-577X}}

%
\authorrunning{J R. Ashford}


\institute{Security, Crime, and Intelligence Innovation Institute, Cardiff University, \email{ashfordjr@cardiff.ac.uk}}

%
\maketitle              
\begin{abstract}
Social media platforms such as Twitter have a fundamental role in facilitating the spread and discussion of ideas online through the concept of retweeting and replying. However, these features also contribute to the spread of mis/disinformation during the vaccine rollout of the COVID-19 pandemic. Using COVID-19 vaccines as a case study, we analyse multiple social network representation derived from three message-based interactions on Twitter (quote retweets, mentions and replies) based upon a set of known anti-vax hashtags and keywords. Each network represents a certain hashtag or keyword which were labelled as ``controversial'' and ``non-controversial'' according to a small group of participants. For each network, we extract a combination of global and local network-based metrics which are used as feature vectors for binary classification. Our results suggest that it is possible to detect controversial from non-controversial terms with high accuracy using simple network-based metrics. Furthermore, these results demonstrate the potential of network representations as language-agnostic models for detecting mis/disinformation at scale, irrespective of content and across multiple social media platforms.

\end{abstract}

\section{Introduction}
Social networks have a fundamental role in the way in which users communicate with one another online. As a result of this, issues such as misinformation begin to emerge due to the size and heavily connected nature of social media platforms \cite{yu2018adversarial,wang2019systematic}. More specifically, since the COVID-19 pandemic, discussions surrounding vaccine usage has attracted highly emotive, positive and negatives view-points which, consequently, increases the potential for misinformation to emerge \cite{hung2020social,monselise2021topics}. This can have far-reaching consequences in an offline setting as ill-informed decisions, as of a result of misinformation, can be a threat to public health \cite{vogel2017viral,wang2019systematic}.

Furthermore, microblogging platforms such as X (known as Twitter in the context of this paper) have further contributed to this issue due to the spread of misinformation surrounding vaccines through the use of message-led interactions (e.g. replies, retweets etc) \cite{sharma2022covid,warner2022vaccine}. What makes Twitter a particularly interesting platform to study is how user's use the platform to communicate with others using three distinct types of message-based interactions. These interactions include \textbf{mentions} (a tweet which contain ``\textit{mentions}'' another person's username), \textbf{replies} (a user responding to another user's tweet) and \textbf{quote retweets} (tweeting another person's tweet with a comment). These pairwise user interactions support different modes of communication and user engagement. 



\begin{figure}[h!]
\centering
\begin{subfigure}[b]{0.3\textwidth}
  \centering
  \includegraphics[scale=0.07]{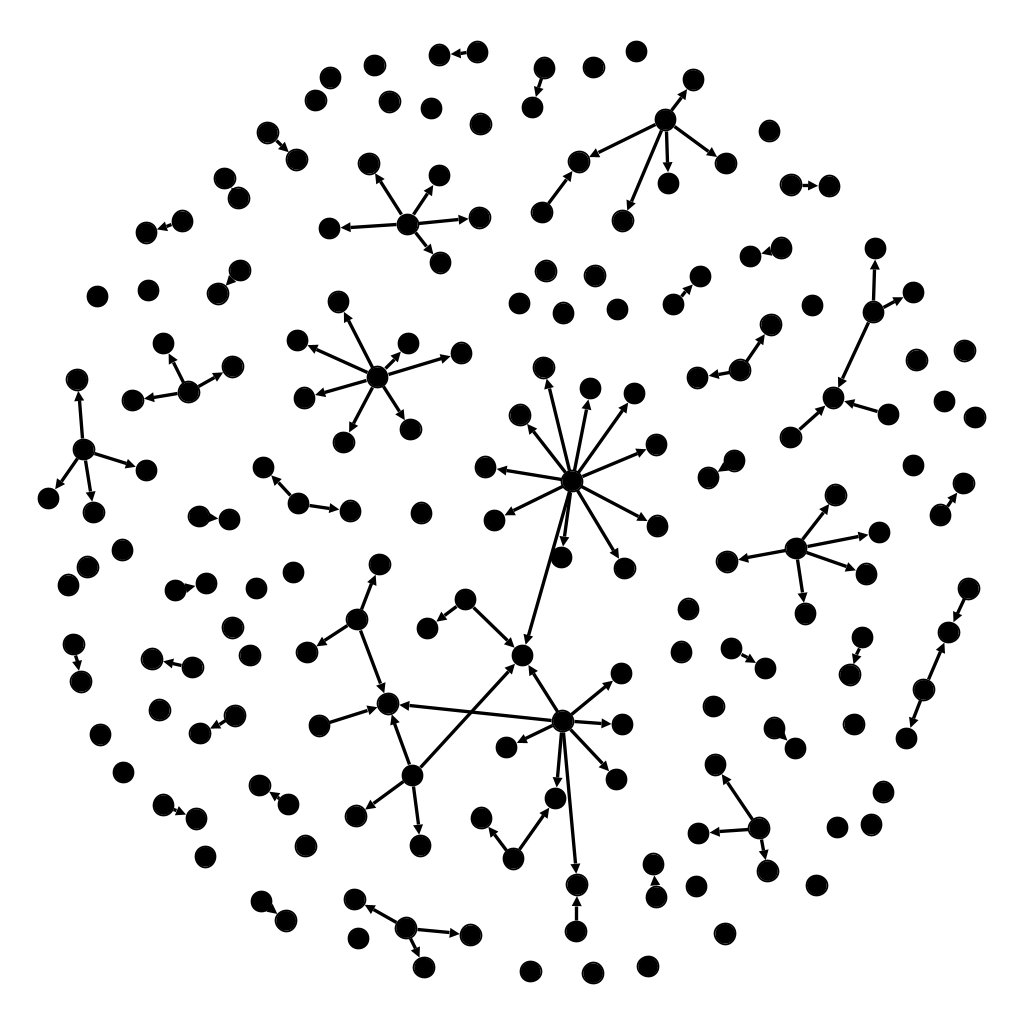}
  \caption{Mention}
  \label{fig:e1}
\end{subfigure}
\begin{subfigure}[b]{0.3\textwidth}
  \centering
  \includegraphics[scale=0.07]{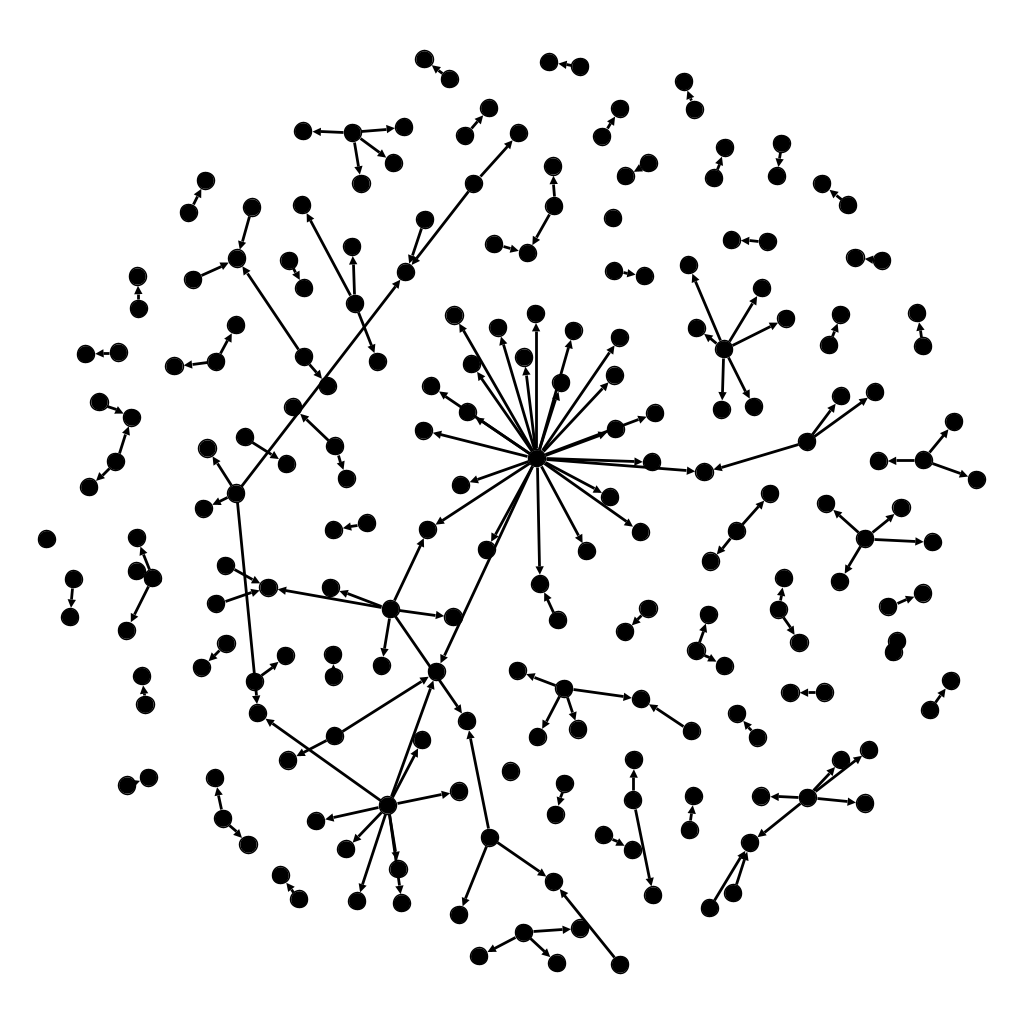}
  \caption{Quote retweet}
  \label{fig:e2}
\end{subfigure}
\begin{subfigure}[b]{0.3\textwidth}
  \centering
  \includegraphics[scale=0.07]{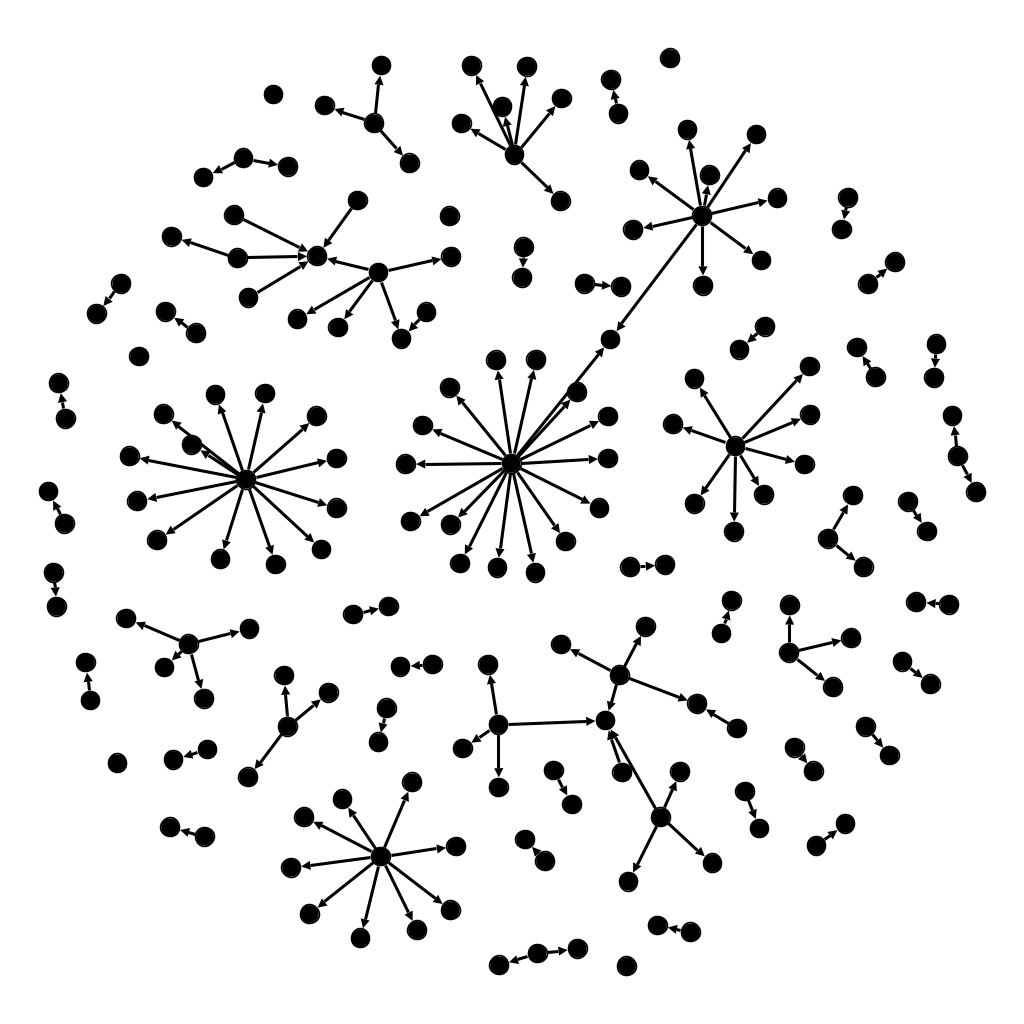}
  \caption{Reply}
  \label{fig:e3}
\end{subfigure}
\caption{Examples of three message-based interaction networks in the form of mentions (left), quote retweets (centre) and replies (right) on Twitter based on tweets which mention \#NoNewNormal.}
\label{fig:example_nonewnormal}
\end{figure}

The three interaction mentioned (mention, reply and quote retweet) are the focus of this paper and are represented in the form of a social network for modelling user activity. An example is shown in Figure \ref{fig:example_nonewnormal}.

In this paper, ``anti-vaccine content'' is defined using a combination of known, crowdsourced and custom COVID-19 anti-vax hashtags and keywords (collectively referred to as ``terms'') used on Twitter which are ranked by a small group of participants ($N=5$). Each term is represented by three individual networks from each interaction type. A term is labelled either ``controversial'' or ``non-controversial'' depending on how it scores on a Likert scale.

We hypothesise that \textit{network metrics and substructures can be used to differentiate between controversial and non-controversial terms using either or a combination of networks constructed from quote retweets, mentions, and replies}. 

In doing so, the aim of this paper is two-fold. Firstly, to understand how network-based features can be used to observe nuances between each network term and interaction type. Little research has been performed in an attempt to study the utility of these three types of interaction (both combined and in isolation) by cross comparison. And secondly, to extract latent signals within each network to differentiate between controversial and non-controversial terms.


\subsection{Background and Related Work}
The ability to detect misinformation and fake news on social media is by no means a novel idea and there has been a wealth of research invested in computational solutions that can identify and detect user accounts responsible for spreading such information in a semi autonomous fashion \cite{shao2016hoaxy,deverna2021covaxxy}. More specifically, misinformation surrounding vaccine usage has been a historical issue which has been exacerbated in recent years due to the COVID-19 pandemic \cite{gallegos2022anti}.

Within the literature, a significant component for predicting anti-vax content relies on the use of natural language processing (NLP) as a solution for analysing textual information. Research has demonstrated that the use of NLP and other methods have the potential to identify tweets containing misinformation \cite{hossain2020covidlies,antypas2021covid}, conspiracies \cite{muric2021covid,li2020constructing} and hate speech \cite{ziems2020racism, pei2020coronavirus}. Within the NLP-based literature, a subset of research focuses specifically on the role of sentiment analysis for identifying both positive and negative view points. Sentiment analysis has been used to understand the various themes and trends surrounding support and opposition towards vaccines \cite{hung2020social,monselise2021topics,kang2017semantic}. The research shows that users promoting anti-vax related content frequently engaged in replies and, overall, were more negative, showing emotions such as rage and sorrow as one of the key themes \cite{miyazaki2021characterizing,chopra2021mining}. Alternatively, a network science-based approach has been used to demonstrate how anti-vax users form echo chambers of polarised communities with other like-minded users by retweeting each other's content \cite{milani2020visual}. 


As of this writing, there is a gap in the literature for using network-based methods exclusively for detecting anti-vax related content on Twitter. The use of social network analysis on Twitter has important implications to this study by considering the utility of the platform's three message-based interactions - retweets, mentions and replies.

To begin, the role of retweet networks have been used to identify communities of like-minded individuals \cite{cherepnalkoski2016retweet} and understand the spread of information \cite{lerman2010information} which, consequently, also includes misinformation \cite{yu2018adversarial}. Retweet networks provide predictive signals for predicting retweeting behaviour \cite{yang2010understanding} and can be used as a reliable proxy for gauging popularity, social capital and friendship formation \cite{recuero2011does, ahn2015structural}.

In addition to this, the use of mention interactions are primarily used to direct a tweet to a user (or group of users) by mentioning their username in their tweet. This type of interaction has been observed in multiple settings, demonstrating that mentions provide utility for predicting links between users and \cite{borge2017opinion,hours2016link} and shares similar structural properties to networks produced using the ``favourite'' button \cite{kato2012network}. Finally, replies can also be treated as a network representation for modelling conversational dynamics. Reply networks have been used to study patterns in multiple Q\&A discussions \cite{rzeszotarski2014anyone}, monitor audience approval or disapproval \cite{molyneux2019political} and to gain followers \cite{rossi2012conversation} in a language-agnostic capacity \cite{bruns2014arab}. Research has demonstrated how replies can be used to validate Dunbar number revealing that much like offline interactions, online interactions can reach a limit to the number of possible interactions that can be preserved \cite{gonccalves2011modeling}. 

Overall, in this paper, we expand upon existing literature by using Twitter as part of a case study to assess the ways in which network-based approaches can be used to model multiple message-based interactions. The literature demonstrates how these interactions have potential to provide predictive utility for detecting controversial and non-controversial networks among various anti-vax topics.

\section{Dataset}
All tweets that are used in this study are a part of the Coronavirus (COVID-19) Tweets Dataset \cite{lamsal2021design}. This dataset serves as a baseline for our analysis as it uses a set of broad, predetermined, generic keywords which are of relevance to the COVID-19 pandemic. In particular, we focus our analysis on a specific window between 9th November and 8th December 2020. This date range refers to the period when the initial vaccines were first approved for use in the United Kingdom thus starting the conversation and reactions around vaccines on Twitter \cite{cotfas2021longest}.

In addition to this, this paper introduces a set of potentially controversial hashtags and keywords (collectively referred to as ``terms''). There are two papers within the supporting literature which provide useful examples of terms which align with disruptive activity. 

The first paper ``\#Scamdemic, \#Plandemic, or \#Scaredemic: What Parler Social Media Platform Tells Us about COVID-19 Vaccine'' uses data originating from Parler \cite{baines2021scamdemic}. While this paper does not focus on Twitter itself, Parler as a platform is well known for its issues regarding echo chambers, filter bubbles and lack of fact-checking \cite{bright2022individuals,alatawi2021survey}.

The second paper of interest (``COVID-19 Vaccine Hesitancy on Social Media: Building a Public Twitter Data Set of Antivaccine Content, Vaccine Misinformation, and Conspiracies'') focuses on attempts to build a set of anti-vax related content by manually collecting keywords which co-occur with previously known / observed anti-vax terms such as \#vaccineskill or \#vaccinedamage \cite{muric2021covid}. Furthermore, their results indicate that 229,041 (12.5\%) of their tweets originated from the UK. This means that the dataset is likely to feature these hashtags, especially within our specified timeframe. As a result, the combination of an applied date range and relevant terms ensures that the tweets used as part of the study are in the scope of vaccine-related content

As well as these two papers, an additional set of terms are used featuring the original keywords / hashtags used as part of the IEEE Coronavirus (COVID-19) Tweets Dataset along with a few terms that were manually selected using the Twitter search tool.

\section{Methodology}

\subsection{Ranking of Terms}\label{m_rot}
To begin, a total of ($N=5$) participants were recruited and were asked to rate each unique term on a Likert scale where each term is scored according to a weight $w \in [0..4]$ based upon the following scale: ``Neutral'' ($w=0$), ``Somewhat Controversial'' ($w=1$), ``Controversial'' ($w=2$), ``Very Controversial'' ($w=3$), ``Highly Controversial'' ($w=4$). The results from each participant are then aggregated to include the total (sum of scores), mean and standard deviation of each score. Each of the terms were then ranked according to mean score. 

\subsection{Hydrating tweets}

In order to retrieve the specific content and metadata (such as the timestamp, reply to, retweet and mention fields) from the Coronavirus (COVID-19) Tweets Dataset, the tweets need to be ``hydrated'' from the original ID where the Twitter API is required to lookup and retrieve the original tweet. In doing so, the process of ``hydrating'' takes the list of IDs provide by the IEEE dataset and transforms them into a set of tweets complete with the information need for subsequent analysis.

\subsection{Network generation}
As mentioned previously, this study focuses on three different types of interaction of interest: quote retweets, mentions and replies. A single network $G_i = (V, E)$ is generated for each interaction type. A node $v_i \in V$ represents a user and the presence of a directed edge $(v_i, v_j) \in E$ indicates an interaction towards another user. For example,  $v_i \rightarrow v_j$, can be interpreted as ``$v_i$ \textit{mentions/replies to/quote retweets from} $v_j$''. 

For each term used in this paper, the three interaction networks generated are conditioned on the presence of the term appearing in the body of a tweet. For example, a subset $t$ of all tweets $t \subset T$ are determined by only focusing on tweets which contain the term ``\#covid19''. For all tweets in this subset, three distinct networks are extracted according to the presence of one of the interactions of interest. Overall, a total of $N=199$ terms are considered focusing on $M=3$ interactions of interest producing a total of $N \times M = 597$ unique networks.

\subsection{Network features}

To evaluate the utility of these network representations, a classification task is used to evaluate how well they perform in predicting controversial terms from non-controversial terms. This is achieved using two sets of network features - global and local network features. 

\subsubsection{Global network features}\label{global_net_f_twitter}
This paper considers the network-based metrics at a global-level by observing how users in each interaction network behave collectively. These include density (the capacity of how many interaction edges occupy the network), reciprocity (the ratio of bidirectional ties in the network), transitivity (the extent to which nodes for transitive edges in a triad), in degree (mean,max,min), out degree (mean,max,min). 

These metrics provide genetic structural properties and also serve as a baseline to determine the predictive utility in comparison to the local network features. Properties such as density, reciprocity and transitivity are fundamental for capturing social traits such as trust, friendships and communities within social networks \cite{liu2011trust,block2015reciprocity,subramani2011density}. These, in turn, can be used to provide predictive signals for differentiating between controversial and non-controversial terms.

\subsubsection{Local network features}\label{local_net_f_twitter}


In addition to global network features, local network features are derived by counting the frequency of all induced subgraphs of a certain size. We described these as local features as they are used to understand the fundamental structure or ``building blocks'' of the network based upon interactions which take place between users. Due to the subgraph isomorphism problem \cite{cook1971complexity}, subgraph counting does not scale well with time and is therefore resource intensive. This means that counting subgraphs grows exponentially with time and that larger subgraphs take much longer to compute. For this reason, we focused on subgraphs with 3 and 4 nodes producing $N=13$ and $N=199$ possible combinations respectively, with a total of $N=212$ subgraphs overall. Each interaction network for a given term produces a vector $V_{G_i}$ where:

\begin{equation}
    V_{G_i} = (v_1, v_2, \dots, v_{212})
\end{equation}

and where $v_i$ represent the frequency of the $i$th subgraph in the set. In addition to this, each of these vectors $V_{G_i}$ are normalised making it possible to compare to networks of different sizes using the following:

\begin{equation}
    V_{G_i} = \frac{1}{\Sigma_{j=1}^{212} v_j} (v_1, v_2, \dots, v_{212})
\end{equation}

As a result, $V_{G_i}$ is used to represent the ratio of subgraph frequencies and provides the basis for discovering under and over-representations of induced subgraphs with respect to other networks. Furthermore, this approach can be used to determine the extent to which interactions and terms share similar structural features.

\section{Results}

\subsection{Discovery of anti-vax terms}
A crowdsourced ranking task was used to discover controversial Twitter terms from non-controversial terms. The combined results from all participants reveal that most terms were labelled as ``Neutral'' which would appear, on average, 37.9\% of the time. The most uncommon label is ``Highly Controversial'' at 8.74\%. 

\begin{table}[h!]
\centering
\begin{tabular}{|l|l|}
\hline
\textbf{Label}         & \textbf{\% of Appearing} \\ \hline
Neutral                & 37.89\%                  \\ \hline
Somewhat Controversial & 20.8\%                   \\ \hline
Controversial          & 19.9\%                   \\ \hline
Very Controversial     & 12.6\%                  \\ \hline
Highly Controversial   & 8.74\%                   \\ \hline
\end{tabular}
\label{label_appearance_tab}
\caption{Full list of labels used within the Likert scale and the probability of appearing}
\end{table}

\begin{figure}
    \centering
    \includegraphics[scale=0.4]{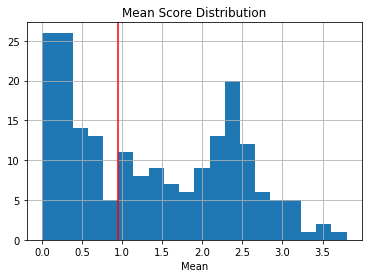}
    \caption{Distribution of mean score for all terms in the set with the threshold $t=0.95$ marked in red}
    \label{fig:mean_score_dist}
\end{figure}

The distribution of mean term score is shown in Figure \ref{fig:mean_score_dist} is used to determine the position of a threshold $t$ as a cut-off point for separating non-controversial and controversial terms. 

Based upon the distribution of mean scores (as shown in Figure \ref{fig:mean_score_dist}) terms are partitioned into the two groups according to a set threshold of $t=0.95$ where values are momentarily reduced before increasing again. 

Terms with an average score exceeding the threshold are considered ``controversial'' and terms lower than the threshold are considered ``non-controversial''. Using this classification technique, a total of ($N=115$) controversial and ($N=84$) non-controversial terms were discovered producing a 58/42 split. 

\subsection{Data Overview}\label{data_overview_tweets}
Using the labels provided as part of the classification task, the data was split into the two sets according to the appropriate label. As a result of computing the global and local metrics, a few additional observations of interest emerged when principal component analysis is performed to determine the spatial relevance of each set of features. These are outlined as follows:

\subsubsection{Global network features}\label{global_overview_tweets}

\begin{figure}[h!]
    \centering
    \includegraphics[scale=0.4]{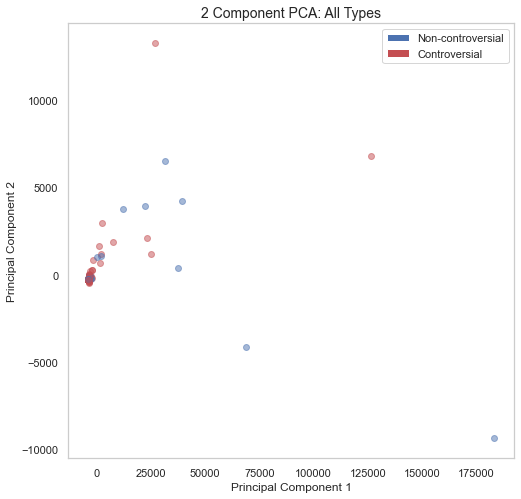}
    \caption{Two-dimensional principal component analysis for all global network features combining reply, mention and quote retweet interactions}
    \label{fig:pca_global_tweets}
\end{figure}

\begin{figure}[h!]
    \centering
    \includegraphics[scale=0.5]{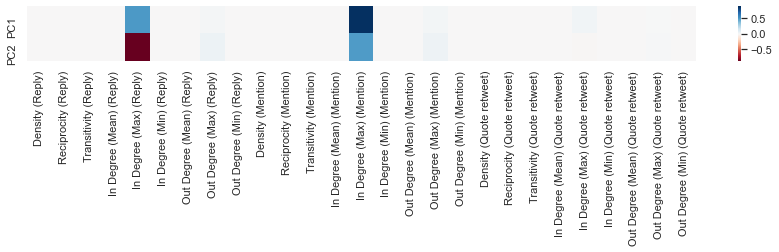}
    \caption{The corresponding eigenvector values for each principal component as shown in Figure \ref{fig:pca_global_tweets}}
    \label{fig:pca_global_tweets_coef}
\end{figure}

Principal component analysis is used to determine the spatial relevance of all global network features, where all three interaction types are combined and cross-compared. These results can be observed in Figure \ref{fig:pca_global_tweets} with supplemented with the corresponding eigenvector values in Figure \ref{fig:pca_global_tweets_coef}.

The PCA scatter plot in Figure \ref{fig:pca_global_tweets} reveals how there are no obvious spacial clustering or patterns which emerge between the two types and provides little spatial utility. The PCA eigenvectors demonstrate how ``In Degree (Max) (Reply)'' and ``In Degree (Max) (Mention)'' appear as the strongest features used for each principal component.

\subsubsection{Local network features}\label{local_overview_tweets}
\begin{figure}[h!]
    \centering
    \includegraphics[scale=0.5]{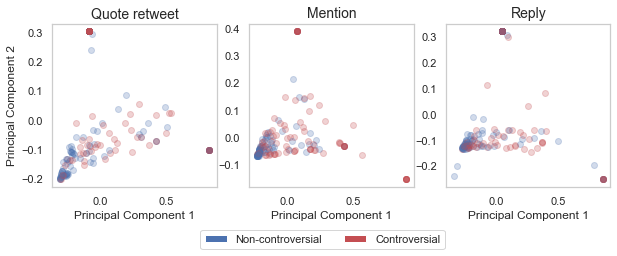}
    \caption{Two-dimensional principal component analysis for all local network features for each interaction type}
    \label{fig:pca_local_tweets}
\end{figure}

\begin{figure}[h!]
    \centering
    \includegraphics[scale=0.4]{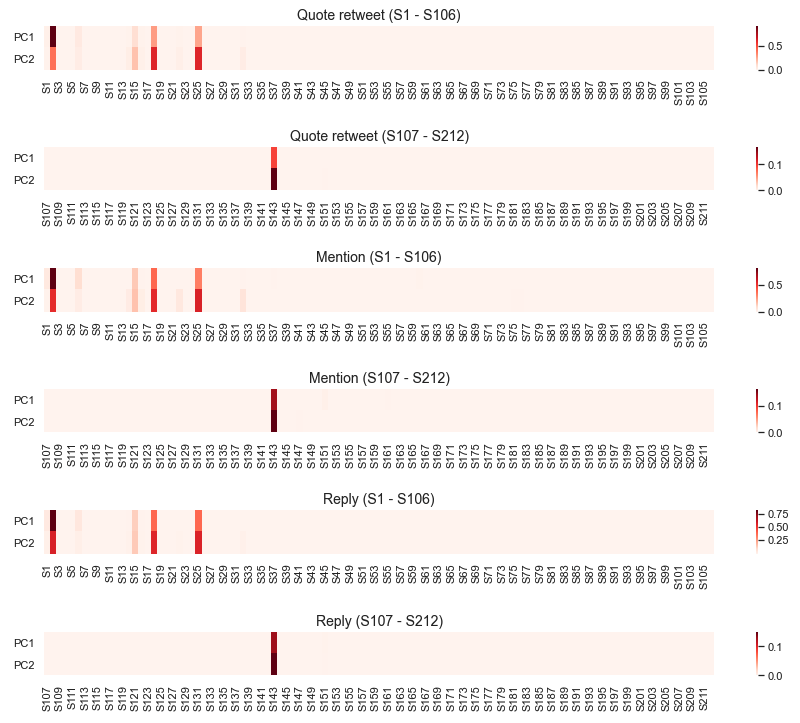}
    \caption{Eigenvector values for each principal component and interaction type as shown in Figure \ref{fig:pca_local_tweets}}
    \label{fig:pca_local_tweets_coef}
\end{figure}

Due to the size of each of the feature vectors ($N=212$ for each type of interaction), PCA is performed to reduce the size of the feature space making it possible visualise the data in two dimensions for each type of interaction. Furthermore, the PCA eigenvectors are used to determine inflectional subgraphs which contribute to the spacial positioning of each feature vector. These results are presented in Figures \ref{fig:pca_local_tweets} and \ref{fig:pca_local_tweets_coef} for the PCA scatter plots and eigenvector values respectively. 

\begin{figure}[h!]
\centering
\begin{subfigure}[b]{0.19\textwidth}
  \centering
  \includegraphics[scale=0.3]{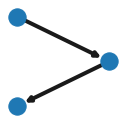}
  \caption{}
  \label{fig:pcs_coef_sg1}
\end{subfigure}
\begin{subfigure}[b]{0.19\textwidth}
  \centering
  \includegraphics[scale=0.3]{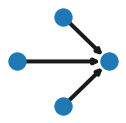}
  \caption{}
  \label{fig:pcs_coef_sg2}
\end{subfigure}
\begin{subfigure}[b]{0.19\textwidth}
  \centering
  \includegraphics[scale=0.3]{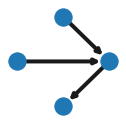}
  \caption{}
  \label{fig:pcs_coef_sg3}
\end{subfigure}
\begin{subfigure}[b]{0.19\textwidth}
  \centering
  \includegraphics[scale=0.3]{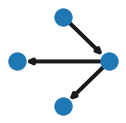}
  \caption{}
  \label{fig:pcs_coef_sg4}
\end{subfigure}
\begin{subfigure}[b]{0.19\textwidth}
  \centering
  \includegraphics[scale=0.3]{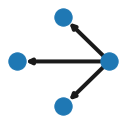}
  \caption{}
  \label{fig:pcs_coef_sg5}
\end{subfigure}
\caption{Subset of subgraphs selected from the PCA eigenvectors for local features used to provide spacial relivance}
\label{fig:pcs_coef_sg}
\end{figure}

The results in Figures \ref{fig:pca_local_tweets} and \ref{fig:pca_local_tweets_coef} suggest that there are no obvious distinction between each of the three interaction types and little indication of clustering potential. Additionally, the coefficients in Figure \ref{fig:pca_local_tweets_coef} reveal that the same set of subgraphs are dominant throughout each of the interaction types. By setting a threshold of $t=0.1$ (determined by observing the distribution of values), across all the eigenvector values, a total of five subgraphs emerged which exceeded this threshold. These include subgraphs \verb|S2|, \verb|S15|, \verb|S18|, \verb|S25| and \verb|S143| which appear consistently across all three interactions and are shown in Figure \ref{fig:pcs_coef_sg} for reference.

\subsection{Classification of controversial terms}\label{class_tweets}
Using the data produced in earlier tasks, we use the feature vectors produced from global and local features (see Sections \ref{local_net_f_twitter} and \ref{global_net_f_twitter}) with the intention of understanding the predictive utility of both global and local network features.


To archive this, we apply binary logistic regression (BLR), support vector machine (SVM) and a random forest classifier (RFC) applied with 10-fold cross-validation. To assess the classification performance, we report the Accuracy, F1 Score, Precision, Recall, Sensitivity, Specificity, Positive predictive value (+PV) and Negative predictive value (-PV).

\subsubsection{Global network features}\label{global_pred_r_tweets}
We train each of the three classification models using the raw metrics defined in Section \ref{global_net_f_twitter}. Our classification results are reported for each type of interaction and combined. The results can be found in Table \ref{global_pred_iso} and \ref{global_pred} for each interaction in isolation and combined respectively.

\begin{table}[h!]
\centering
\begin{tabular}{l|lll|lll|lll|}
\cline{2-10}
\textbf{}                                      & \multicolumn{3}{l|}{\textbf{Mention}} & \multicolumn{3}{l|}{\textbf{Quote retweet}} & \multicolumn{3}{l|}{\textbf{Reply}} \\ \hline
\multicolumn{1}{|l|}{\textbf{Classifier}}      & BLR      & SVM      & RFC             & BLR        & SVM       & RFC                & BLR     & SVM     & RFC             \\ \hline
\multicolumn{1}{|l|}{\textbf{Accuracy}}        & 0.744    & 0.619    & \textbf{0.81}   & 0.68       & 0.609     & \textbf{0.746}     & 0.698   & 0.598   & \textbf{0.716}  \\ \cline{1-1}
\multicolumn{1}{|l|}{\textbf{F1 Score}}        & 0.792    & 0.733    & 0.83            & 0.748      & 0.718     & 0.782              & 0.758   & 0.724   & 0.742           \\ \cline{1-1}
\multicolumn{1}{|l|}{\textbf{Precision}}       & 0.695    & 0.583    & 0.788           & 0.635      & 0.575     & 0.706              & 0.656   & 0.567   & 0.711           \\ \cline{1-1}
\multicolumn{1}{|l|}{\textbf{Recall}}          & 0.921    & 0.989    & 0.876           & 0.909      & 0.955     & 0.875              & 0.899   & 1.0     & 0.775           \\ \cline{1-1}
\multicolumn{1}{|l|}{\textbf{Sensitivity}}     & 0.544    & 0.203    & 0.734           & 0.432      & 0.235     & 0.605              & 0.475   & 0.15    & 0.65            \\ \cline{1-1}
\multicolumn{1}{|l|}{\textbf{Specificity}}     & 0.921    & 0.989    & 0.876           & 0.909      & 0.955     & 0.875              & 0.899   & 1.0     & 0.775           \\ \cline{1-1}
\multicolumn{1}{|l|}{\textbf{+PV}} & 0.695    & 0.583    & 0.788           & 0.635      & 0.575     & 0.706              & 0.656   & 0.567   & 0.711           \\ \cline{1-1}
\multicolumn{1}{|l|}{\textbf{-PV}} & 0.86     & 0.941    & 0.841           & 0.814      & 0.826     & 0.817              & 0.809   & 1.0     & 0.722           \\ \hline
\end{tabular}
\caption{Complete classification results for all three interaction types using global features reporting the performance for each classifier. The best performing classifier is highlighted in bold}
\label{global_pred_iso}
\end{table}

The results in Table \ref{global_pred_iso} show the RFC outperforms both SVM and BLR consistently across each type of interaction using global features. As highlighted in Table \ref{global_pred_iso}, mention interactions combined with an RFC produces the best performing classifier with an accuracy of $p=0.81$. This is then followed by quote retweets with an accuracy of $p=0.746$ and finally, reply with an accuracy of $p=0.719$

In addition to this, we perform a separate classification task whereby all global network features for each interaction are combined into a single feature vector. All interaction types are combined to assess whether this has an impact on classification performance. These results can be found in Table \ref{global_pred}.

\begin{table}[h!]
\centering
\begin{tabular}{|l|lll|}
\hline
\textbf{Classifier}      & \multicolumn{1}{l|}{\textbf{BLR}} & \multicolumn{1}{l|}{\textbf{SVM}} & \textbf{RFC}   \\ \hline
\textbf{Accuracy}        & 0.847                             & 0.852                             & \textbf{0.886} \\ \cline{1-1}
\textbf{F1 Score}        & 0.914                             & 0.918                             & 0.937          \\ \cline{1-1}
\textbf{Precision}       & 0.873                             & 0.869                             & 0.887          \\ \cline{1-1}
\textbf{Recall}          & 0.96                              & 0.973                             & 0.993          \\ \cline{1-1}
\textbf{Sensitivity}     & 0.192                             & 0.154                             & 0.269          \\ \cline{1-1}
\textbf{Specificity}     & 0.96                              & 0.973                             & 0.993          \\ \cline{1-1}
\textbf{+ Predict Value} & 0.873                             & 0.869                             & 0.887          \\ \cline{1-1}
\textbf{- Predict Value} & 0.455                             & 0.5                               & 0.875          \\ \hline
\end{tabular}
\caption{Classification results combining all three interaction types using global features. The best performing classifier is highlighted in bold}
\label{global_pred}
\end{table}

The results in Table \ref{global_pred} indicate that by combining all interaction types, the accuracy of the RFC increases to around $p=0.886$. This produces a performance gain of approximately 9.3\% compared with the mention RFC accuracy - the best performing classifier of all interaction types.

\subsubsection{Local network features}
Following the same format as Section \ref{global_pred_r_tweets}, we train the three same classifiers (RFC, BLR and SVM) using local network features exclusively. The classification results for each interaction type can be found in Table \ref{local_pred_iso}.

\begin{table}[h!]
\centering
\begin{tabular}{l|lll|lll|lll|}
\cline{2-10}
\textbf{}                                      & \multicolumn{3}{l|}{\textbf{Mention}} & \multicolumn{3}{l|}{\textbf{Quote retweet}} & \multicolumn{3}{l|}{\textbf{Reply}} \\ \hline
\multicolumn{1}{|l|}{\textbf{Classifier}}      & BLR      & SVM     & RFC              & BLR        & SVM       & RFC                & BLR     & SVM              & RFC    \\ \hline
\multicolumn{1}{|l|}{\textbf{Accuracy}}        & 0.72     & 0.732   & \textbf{0.786}   & 0.663      & 0.663     & \textbf{0.704}     & 0.663   & \textbf{0.692}   & 0.615  \\ \cline{1-1}
\multicolumn{1}{|l|}{\textbf{F1 Score}}        & 0.715    & 0.731   & 0.8              & 0.667      & 0.692     & 0.747              & 0.674   & 0.711            & 0.677  \\ \cline{1-1}
\multicolumn{1}{|l|}{\textbf{Precision}}       & 0.776    & 0.782   & 0.791            & 0.687      & 0.66      & 0.673              & 0.686   & 0.703            & 0.607  \\ \cline{1-1}
\multicolumn{1}{|l|}{\textbf{Recall}}          & 0.663    & 0.685   & 0.809            & 0.648      & 0.727     & 0.841              & 0.663   & 0.719            & 0.764  \\ \cline{1-1}
\multicolumn{1}{|l|}{\textbf{Sensitivity}}     & 0.785    & 0.785   & 0.759            & 0.679      & 0.593     & 0.556              & 0.662   & 0.662            & 0.45   \\ \cline{1-1}
\multicolumn{1}{|l|}{\textbf{Specificity}}     & 0.663    & 0.685   & 0.809            & 0.648      & 0.727     & 0.841              & 0.663   & 0.719            & 0.764  \\ \cline{1-1}
\multicolumn{1}{|l|}{\textbf{+PV}} & 0.776    & 0.782   & 0.791            & 0.687      & 0.66      & 0.673              & 0.686   & 0.703            & 0.607  \\ \cline{1-1}
\multicolumn{1}{|l|}{\textbf{-PV}} & 0.674    & 0.689   & 0.779            & 0.64       & 0.667     & 0.763              & 0.639   & 0.679            & 0.632  \\ \hline
\end{tabular}
\caption{Complete classification results for all three interaction types using local features reporting the performance for each classifier. The best performing classifier is highlighted in bold}
\label{local_pred_iso}
\end{table}

Much like Section \ref{global_pred_r_tweets}, RFC is the highest performing classifier for each interaction type except for reply interactions where SVM performs the best. Similarly, mention interactions outperform quote retweets and replies with respect to predictive utility and performance. Each of the feature vectors used as part of the classification task in Table \ref{local_pred_iso} are combined and used in a separate classification task with the results shown in Table \ref{local_pred}.

\begin{table}[h!]
\centering
\begin{tabular}{|l|lll|}
\hline
\textbf{Classifier}      & \multicolumn{1}{l|}{\textbf{BLR}} & \multicolumn{1}{l|}{\textbf{SVM}} & \textbf{RFC} \\ \hline
\textbf{Accuracy}        & 0.852                             & 0.852                             & \textbf{0.858 }       \\ \cline{1-1}
\textbf{F1 Score}        & 0.92                              & 0.92                              & 0.922        \\ \cline{1-1}
\textbf{Precision}       & 0.852                             & 0.852                             & 0.865        \\ \cline{1-1}
\textbf{Recall}          & 1.0                               & 1.0                               & 0.987        \\ \cline{1-1}
\textbf{Sensitivity}     & 0.0                               & 0.0                               & 0.115        \\ \cline{1-1}
\textbf{Specificity}     & 1.0                               & 1.0                               & 0.987        \\ \cline{1-1}
\textbf{+ Predict Value} & 0.852                             & 0.852                             & 0.865        \\ \cline{1-1}
\textbf{- Predict Value} & -                                 & -                                 & 0.6          \\ \hline
\end{tabular}
\caption{Classification results combining all three interaction types using local features. The best performing classifier is highlighted in bold}
\label{local_pred}
\end{table}

The results in Table \ref{local_pred} show that RFC is the best performing classifier when all interactions are combined using local features. The accuracy of the RFC mode increased to $p=0.858$ which, in turn, produces a performance gain of 9.16\% compared with the best performing result in Table \ref{local_pred_iso}.

\section{Discussion}
Both the data overview (see Section \ref{data_overview_tweets}) and classification results (see Section \ref{class_tweets}) provide meaningful insights on the utility of using network representations for predicting controversial and non-controversial terms. As a result, a number of key observations are explored and disused further in this section which relate to the utility of both global and local network representations and their prediction performance. These are discussed as follows:

As described in Section \ref{local_overview_tweets}, the PCA eigenvectors shown in Figure \ref{fig:pca_local_tweets_coef} reveals five subgraphs (see Figure \ref{fig:pcs_coef_sg} for subgraphs \verb|S2|, \verb|S15|, \verb|S18|, \verb|S25| and \verb|S143|) which exceed an arbitrary threshold of $t=0.1$. These subgraphs reflect those that resemble a tree-like structure where all interactions are centred around one node (similar to the previous study). This potentially correlates with features such as the maximum in degree of a network where many nodes are directed towards a single ``central'' node. This is evident based upon the features which emerged in the PCA eigenvectors for global features in Figure \ref{fig:pca_global_tweets_coef} where the maximum in degree is dominant.


The use of global features for predicting controversial and non-controversial networks reveals rather promising results. The use of a RFC (Random Forest Classifier) consistently outperforms alternative classifiers across each of the three interaction types. The results in Table \ref{global_pred_iso} show that mention interactions can best differentiate between the two types with a classification accuracy of $p=0.81$. Based open the eigenvector values in Figure \ref{fig:pca_global_tweets_coef}, it is possible to speculate that the maximum in degree for mention interactions (one of the most dominant features) provide the best spatial distribution of networks in order to separate the two groups. As a result of combining all three interactions, it is also reasonable to imply that improved accuracy of $p=0.886$ is due to the maximum in degree for replies also being a dominant feature in the PCA eigenvectors in Figure \ref{fig:pca_global_tweets_coef}. By using local features for differentiating between controversial and non-controversial networks a similar trend can be observed compared with global features. The results are fairly consistent across each of the interactions which is unsurprising considering that the PCA eigenvectors in Figure \ref{fig:pca_local_tweets_coef} for each interaction are almost all identical. Similarly, mention interactions provide the best result with respect to accuracy ($p=0.786$) which is improved to $p=0.858$ when all interactions are combined.

\section{Conclusion}

Using COVID-19 anti-vaccine content as a case study, the insights gained from this investigation provide meaningful insights towards understanding the utility of using social network representations for differentiating between controversial and non-controversial. In particular, this paper satisfies the hypothesis by using a prediction task to show how three message-based user interactions (quote retweets, mentions and replies) provide network-based representations to understand how users behave collectively based upon their underlying network substructures and metrics.

The results of this paper provide evidence that simple graph-based metrics such as in/out degree, density, reciprocity and transitivity are sufficient for differentiating between controversial and non-controversial networks. By combining all three interactions into one, it is clear that global features adequately capture the nuances between each of the networks using relatively few features - contrary to subgraph ``local'' approach. As well as providing a performance gain, the set of global features used in this study are relatively easy to calculate and can almost be done in near real-time.


The implications of this paper impact how we are to consider using quote retweets, mentions, and replies (or a combination of the three) in future work. The research featured in this paper clearly demonstrate that it is possible to differentiate between different networks using interactions derived from human behaviour exclusively. The clear advantage of this is that little textual analysis (e.g NLP) is needed, making it possible to replicate results using a non-English speaking corpus. Furthermore, these results have applications in content moderation, whereby moderators can use similar techniques to identify the presence of controversial content using simple network-based metrics. Finally, we emphasise that these techniques are transferable and can be used for a wide range of scenarios for detecting disruptive activity more broadly due to widespread adoption of features such as replying and sharing - fundamental user-to-user interactions relevant to almost all social media platforms.

\bibliography{main}

\begin{thebibliography}{10}

\bibitem{ahn2015structural}
Hyerim Ahn and Ji-Hong Park.
\newblock The structural effects of sharing function on twitter networks: Focusing on the retweet function.
\newblock {\em Journal of Information Science}, 41(3):354--365, 2015.

\bibitem{alatawi2021survey}
Faisal Alatawi, Lu~Cheng, Anique Tahir, Mansooreh Karami, Bohan Jiang, Tyler Black, and Huan Liu.
\newblock A survey on echo chambers on social media: Description, detection and mitigation.
\newblock {\em arXiv preprint arXiv:2112.05084}, 2021.

\bibitem{antypas2021covid}
Dimosthenis Antypas, Jose Camacho-Collados, Alun Preece, and David Rogers.
\newblock Covid-19 and misinformation: A large-scale lexical analysis on twitter.
\newblock In {\em Proceedings of the 59th Annual Meeting of the Association for Computational Linguistics and the 11th International Joint Conference on Natural Language Processing: Student Research Workshop}, pages 119--126, 2021.

\bibitem{baines2021scamdemic}
Annalise Baines, Muhammad Ittefaq, and Mauryne Abwao.
\newblock \# scamdemic,\# plandemic, or\# scaredemic: what parler social media platform tells us about covid-19 vaccine.
\newblock {\em Vaccines}, 9(5):421, 2021.

\bibitem{block2015reciprocity}
Per Block.
\newblock Reciprocity, transitivity, and the mysterious three-cycle.
\newblock {\em Social Networks}, 40:163--173, 2015.

\bibitem{borge2017opinion}
Rosa Borge~Bravo and Marc Esteve Del~Valle.
\newblock Opinion leadership in parliamentary twitter networks: A matter of layers of interaction?
\newblock {\em Journal of Information Technology \& Politics}, 14(3):263--276, 2017.

\bibitem{bright2022individuals}
Jonathan Bright, Nahema Marchal, Bharath Ganesh, and Stevan Rudinac.
\newblock How do individuals in a radical echo chamber react to opposing views? evidence from a content analysis of stormfront.
\newblock {\em Human Communication Research}, 48(1):116--145, 2022.

\bibitem{bruns2014arab}
Axel Bruns, Tim Highfield, and Jean Burgess.
\newblock The arab spring and its social media audiences: English and arabic twitter users and their networks.
\newblock In {\em Cyberactivism on the participatory web}, pages 96--128. Routledge, 2014.

\bibitem{cherepnalkoski2016retweet}
Darko Cherepnalkoski and Igor Mozeti{\v{c}}.
\newblock Retweet networks of the european parliament: Evaluation of the community structure.
\newblock {\em Applied network science}, 1(1):1--20, 2016.

\bibitem{chopra2021mining}
Harshita Chopra, Aniket Vashishtha, Ridam Pal, Ananya Tyagi, Tavpritesh Sethi, et~al.
\newblock Mining trends of covid-19 vaccine beliefs on twitter with lexical embeddings.
\newblock {\em arXiv preprint arXiv:2104.01131}, 2021.

\bibitem{cook1971complexity}
Stephen~A Cook.
\newblock The complexity of theorem-proving procedures.
\newblock In {\em Proceedings of the third annual ACM symposium on Theory of computing}, pages 151--158, 1971.

\bibitem{cotfas2021longest}
Liviu-Adrian Cotfas, Camelia Delcea, Ioan Roxin, Corina Ioan{\u{a}}{\c{s}}, Dana~Simona Gherai, and Federico Tajariol.
\newblock The longest month: analyzing covid-19 vaccination opinions dynamics from tweets in the month following the first vaccine announcement.
\newblock {\em Ieee Access}, 9:33203--33223, 2021.

\bibitem{deverna2021covaxxy}
Matthew~R DeVerna, Francesco Pierri, Bao~Tran Truong, John Bollenbacher, David Axelrod, Niklas Loynes, Christopher Torres-Lugo, Kai-Cheng Yang, Filippo Menczer, and John Bryden.
\newblock Covaxxy: A collection of english-language twitter posts about covid-19 vaccines.
\newblock In {\em ICWSM}, pages 992--999, 2021.

\bibitem{gallegos2022anti}
Miguel Gallegos, Viviane de~Castro~Pecanha, and Tom{\'a}s Caycho-Rodr{\'\i}guez.
\newblock Anti-vax: the history of a scientific problem.
\newblock {\em Journal of Public Health}, 2022.

\bibitem{gonccalves2011modeling}
Bruno Gon{\c{c}}alves, Nicola Perra, and Alessandro Vespignani.
\newblock Modeling users' activity on twitter networks: Validation of dunbar's number.
\newblock {\em PloS one}, 6(8):e22656, 2011.

\bibitem{hossain2020covidlies}
Tamanna Hossain, Robert~L Logan~IV, Arjuna Ugarte, Yoshitomo Matsubara, Sean Young, and Sameer Singh.
\newblock Covidlies: Detecting covid-19 misinformation on social media.
\newblock 2020.

\bibitem{hours2016link}
Hadrien Hours, Eric Fleury, and M{\'a}rton Karsai.
\newblock Link prediction in the twitter mention network: impacts of local structure and similarity of interest.
\newblock In {\em 2016 IEEE 16th International Conference on Data Mining Workshops (ICDMW)}, pages 454--461. IEEE, 2016.

\bibitem{hung2020social}
Man Hung, Evelyn Lauren, Eric~S Hon, Wendy~C Birmingham, Julie Xu, Sharon Su, Shirley~D Hon, Jungweon Park, Peter Dang, Martin~S Lipsky, et~al.
\newblock Social network analysis of covid-19 sentiments: Application of artificial intelligence.
\newblock {\em Journal of medical Internet research}, 22(8):e22590, 2020.

\bibitem{kang2017semantic}
Gloria~J Kang, Sinclair~R Ewing-Nelson, Lauren Mackey, James~T Schlitt, Achla Marathe, Kaja~M Abbas, and Samarth Swarup.
\newblock Semantic network analysis of vaccine sentiment in online social media.
\newblock {\em Vaccine}, 35(29):3621--3638, 2017.

\bibitem{kato2012network}
Shoko Kato, Akihiro Koide, Takayasu Fushimi, Kazumi Saito, and Hiroshi Motoda.
\newblock Network analysis of three twitter functions: favorite, follow and mention.
\newblock In {\em Pacific Rim Knowledge Acquisition Workshop}, pages 298--312. Springer, 2012.

\bibitem{lamsal2021design}
Rabindra Lamsal.
\newblock Design and analysis of a large-scale covid-19 tweets dataset.
\newblock {\em Applied Intelligence}, 51(5):2790--2804, 2021.

\bibitem{lerman2010information}
Kristina Lerman and Rumi Ghosh.
\newblock Information contagion: An empirical study of the spread of news on digg and twitter social networks.
\newblock In {\em Fourth international AAAI conference on weblogs and social media}, 2010.

\bibitem{li2020constructing}
Yachao Li, Sylvia Twersky, Kelsey Ignace, Mei Zhao, Radhika Purandare, Breeda Bennett-Jones, and Scott~R Weaver.
\newblock Constructing and communicating covid-19 stigma on twitter: a content analysis of tweets during the early stage of the covid-19 outbreak.
\newblock {\em International Journal of Environmental Research and Public Health}, 17(18):6847, 2020.

\bibitem{liu2011trust}
Guanfeng Liu, Yan Wang, and Mehmet~A Orgun.
\newblock Trust transitivity in complex social networks.
\newblock In {\em twenty-fifth AAAI conference on artificial intelligence}, 2011.

\bibitem{milani2020visual}
Elena Milani, Emma Weitkamp, and Peter Webb.
\newblock The visual vaccine debate on twitter: A social network analysis.
\newblock {\em Media and Communication}, 8(2):364--375, 2020.

\bibitem{miyazaki2021characterizing}
Kunihiro Miyazaki, Takayuki Uchiba, Kenji Tanaka, and Kazutoshi Sasahara.
\newblock Characterizing the anti-vaxxers’ reply behavior on social media.
\newblock In {\em IEEE/WIC/ACM International Conference on Web Intelligence and Intelligent Agent Technology}, pages 83--89, 2021.

\bibitem{molyneux2019political}
Logan Molyneux and Rachel~R Mour{\~a}o.
\newblock Political journalists’ normalization of twitter: Interaction and new affordances.
\newblock {\em Journalism Studies}, 20(2):248--266, 2019.

\bibitem{monselise2021topics}
Michal Monselise, Chia-Hsuan Chang, Gustavo Ferreira, Rita Yang, Christopher~C Yang, et~al.
\newblock Topics and sentiments of public concerns regarding covid-19 vaccines: social media trend analysis.
\newblock {\em Journal of Medical Internet Research}, 23(10):e30765, 2021.

\bibitem{muric2021covid}
Goran Muric, Yusong Wu, Emilio Ferrara, et~al.
\newblock Covid-19 vaccine hesitancy on social media: building a public twitter data set of antivaccine content, vaccine misinformation, and conspiracies.
\newblock {\em JMIR public health and surveillance}, 7(11):e30642, 2021.

\bibitem{pei2020coronavirus}
Xin Pei and Deval Mehta.
\newblock \# coronavirus or\# chinesevirus?!: Understanding the negative sentiment reflected in tweets with racist hashtags across the development of covid-19.
\newblock {\em arXiv preprint arXiv:2005.08224}, 2020.

\bibitem{recuero2011does}
Raquel Recuero, Ricardo Araujo, and Gabriela Zago.
\newblock How does social capital affect retweets?
\newblock In {\em Fifth International AAAI Conference on Weblogs and Social Media}, 2011.

\bibitem{rossi2012conversation}
Luca Rossi and Matteo Magnani.
\newblock Conversation practices and network structure in twitter.
\newblock In {\em Proceedings of the International AAAI Conference on Web and Social Media}, volume~6, pages 563--566, 2012.

\bibitem{rzeszotarski2014anyone}
Jeffrey~M Rzeszotarski, Emma~S Spiro, Jorge~Nathan Matias, Andr{\'e}s Monroy-Hern{\'a}ndez, and Meredith~Ringel Morris.
\newblock Is anyone out there? unpacking q\&a hashtags on twitter.
\newblock In {\em Proceedings of the SIGCHI Conference on Human Factors in Computing Systems}, pages 2755--2758, 2014.

\bibitem{shao2016hoaxy}
Chengcheng Shao, Giovanni~Luca Ciampaglia, Alessandro Flammini, and Filippo Menczer.
\newblock Hoaxy: A platform for tracking online misinformation.
\newblock In {\em Proceedings of the 25th international conference companion on world wide web}, pages 745--750, 2016.

\bibitem{sharma2022covid}
Karishma Sharma, Yizhou Zhang, and Yan Liu.
\newblock Covid-19 vaccine misinformation campaigns and social media narratives.
\newblock In {\em Proceedings of the International AAAI Conference on Web and Social Media}, volume~16, pages 920--931, 2022.

\bibitem{subramani2011density}
Kumar Subramani, Alexander Velkov, Irene Ntoutsi, Peer Kroger, and Hans-Peter Kriegel.
\newblock Density-based community detection in social networks.
\newblock In {\em 2011 IEEE 5th International Conference on Internet Multimedia Systems Architecture and Application}, pages 1--8. IEEE, 2011.

\bibitem{vogel2017viral}
Lauren Vogel.
\newblock Viral misinformation threatens public health, 2017.

\bibitem{wang2019systematic}
Yuxi Wang, Martin McKee, Aleksandra Torbica, and David Stuckler.
\newblock Systematic literature review on the spread of health-related misinformation on social media.
\newblock {\em Social science \& medicine}, 240:112552, 2019.

\bibitem{warner2022vaccine}
Echo~L Warner, Juliana~L Barbati, Kaylin~L Duncan, Kun Yan, and Stephen~A Rains.
\newblock Vaccine misinformation types and properties in russian troll tweets.
\newblock {\em Vaccine}, 40(6):953--960, 2022.

\bibitem{yang2010understanding}
Zi~Yang, Jingyi Guo, Keke Cai, Jie Tang, Juanzi Li, Li~Zhang, and Zhong Su.
\newblock Understanding retweeting behaviors in social networks.
\newblock In {\em Proceedings of the 19th ACM international conference on Information and knowledge management}, pages 1633--1636, 2010.

\bibitem{yu2018adversarial}
Sixie Yu, Yevgeniy Vorobeychik, and Scott Alfeld.
\newblock Adversarial classification on social networks.
\newblock {\em arXiv preprint arXiv:1801.08159}, 2018.

\bibitem{ziems2020racism}
Caleb Ziems, Bing He, Sandeep Soni, and Srijan Kumar.
\newblock Racism is a virus: Anti-asian hate and counterhate in social media during the covid-19 crisis.
\newblock {\em arXiv preprint arXiv:2005.12423}, 2020.

\end{thebibliography}

\end{document}